# Perspectives for self-driving labs in synthetic biology

*Hector Garcia Martin[1,2,3,4,*], Tijana Radivojevic[1,2,3], Jeremy Zucker[16], Kristofer Bouchard[1,5,6], Jess Sustarich[3,15], Sean Peisert[5,8], Dan Arnold[7], Nathan Hillson[1,2,3], Gyorgy Babnigg[2,13], Jose Manuel Marti[1,2,3,9], Christopher J. Mungall[1], Gregg T. Beckham[2,17], Lucas Waldburger[18], James Carothers[12], ShivShankar Sundaram[10,11], Deb Agarwal[5], Blake A. Simmons[1,2,3], Tyler Backman[1,3], Deepanwita Banerjee[1,3], Deepti Tanjore[1,2,14], Lavanya Ramakrishnan[5], Anup Singh[3,10]*

[1] Lawrence Berkeley National Laboratory, Biological Systems and Engineering Division, Berkeley, CA, United States
[2] Department of Energy, Agile BioFoundry, Emeryville, CA, United States
[3] Joint BioEnergy Institute, Emeryville, CA, United States
[4] BCAM, Basque Center for Applied Mathematics, Bilbao, Spain
[5] Lawrence Berkeley National Laboratory, Scientific Data Division, Berkeley, CA, United States
[6] Helen Wills Neuroscience Institute and Redwood Center for Theoretical Neuroscience, Berkeley, CA, United States
[7] Lawrence Berkeley National Laboratory, Energy Storage and Distributed Resources Division, Berkeley, CA, United States
[8] University of California, Davis, Department of Computer Science, Davis, CA, United States
[9] Global Security Computing Applications Division, Lawrence Livermore National Laboratory, Livermore, CA, United States
[10] Engineering Directorate, Lawrence Livermore National Laboratory, Livermore, CA, United States
[11] Center for Bioengineering, Lawrence Livermore National Laboratory, Livermore, CA, United States
[12] Department of Chemical Engineering, Molecular Engineering & Sciences Institute and Center for Synthetic Biology, University of Washington, Seattle, WA, United States
[13] Biosciences Division, Argonne National Laboratory, Argonne, IL, United States
[14] Advanced Biofuels and Bioproducts Process Development Unit, Lawrence Berkeley National Laboratory, Berkeley, CA, United States
[15] Biomaterials and Biomanufacturing Division, Sandia National Laboratories, Livermore, CA, United States
[16] Earth and Biological Sciences Division, Pacific Northwest National Laboratories, Richland, WA, United States
[17] Resources and Enabling Sciences Center, National Renewable Energy Laboratory, Golden CO 80401 USA
[18] Department of Bioengineering, University of California, Berkeley, CA, United States

*Correspondence:
Hector Garcia Martin
hgmartin@lbl.gov

*Abstract*

Self-driving labs (SDLs) combine fully automated experiments with artificial intelligence (AI) that decides the next set of experiments. Taken to their ultimate expression, SDLs could usher a new paradigm of scientific research, where the world is probed, interpreted, and explained by machines for human benefit. While there are functioning SDLs in the fields of chemistry and materials science, we contend that synthetic biology provides a unique opportunity since the genome provides a single target for affecting the incredibly wide repertoire of biological cell behavior. However, the level of investment required for the creation of biological SDLs is only warranted if directed towards solving difficult and enabling biological questions. Here, we discuss challenges and opportunities in creating SDLs for synthetic biology.

# What is a self-driving lab?

Self-driving labs (SDLs), or autonomous experimentation, combine robotics for automated experiments and data collection, with Artificial Intelligence (AI) systems that use these data to recommend follow-up experiments [1–3] (Fig. 1). These recommendations potentially involve not just the conditions and parts to be used for the next experiment, but also which underlying hypothesis to test.

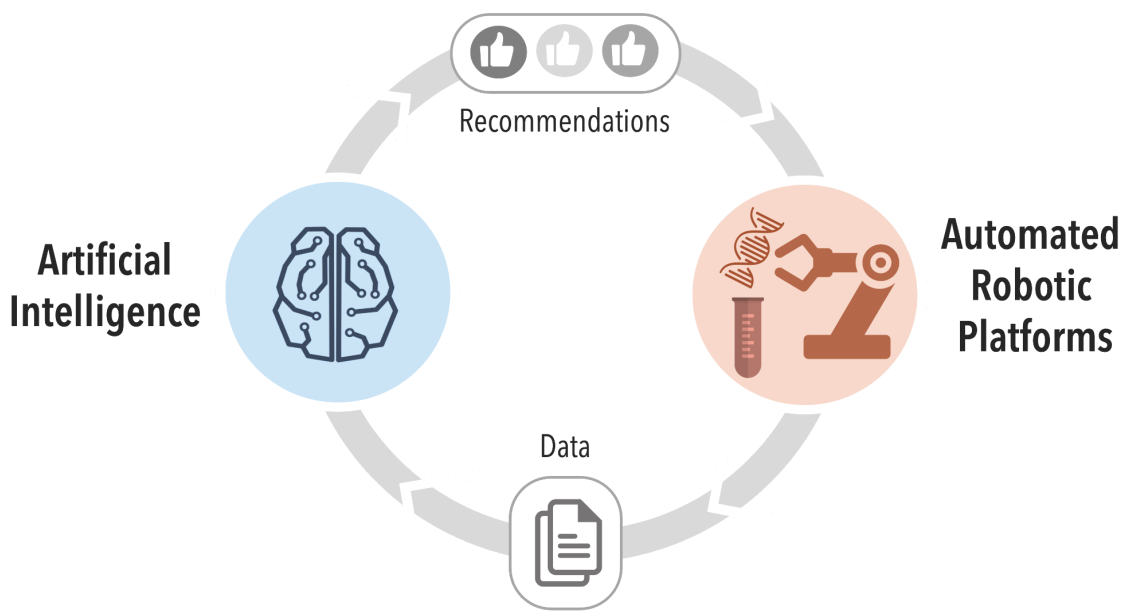

**Fig. 1**: **Self-driving labs (SDLs)** combine automated robotic platforms and data collection with AI that processes these data to decide the next set of experiments to perform and, potentially, which hypotheses and theories to test.

The SDL concept requires full autonomy from humans. A partially automated system, or one that requires human intervention to finish the cycle of experimentation/planning is not, rigorously speaking, a SDL. Full automation for the cycle is not a whimsical requirement, but rather enables the full potential of SDLs. Completely automated systems can reach duty cycles (e.g., 24/7/365 operation), experiment-to-experiment reproducibility, and efficiency that are unattainable by humans. Furthermore, they are potentially linearly scalable (e.g., simply acquire more copies of the equipment), and, as a consequence, can produce large amounts of high-quality data and metadata. Such large volumes of high-quality data can make AI systems particularly effective and insightful: Artificial Neural Networks (ANNs), for example, are known to be most effective once a certain threshold of training data is available. These benefits will be critical to produce improved scientific understanding, and significantly decrease time to desired bio-products.

However crucial, the requirement of full autonomy may be too stringent for the current state of technology, so it is useful to consider intermediate steps toward the realization of SDLs. Similar considerations have moved the car industry to entertain the concept of "Degrees of autonomy" in self-driving cars. For this reason, a similar set of "Autonomy levels" have been proposed to both describe the current technological capabilities and incentivize the gradual development into fully autonomous systems [4] (see Fig. 2). In practical terms, systems displaying an autonomy level of three or above can be considered SDLs, since they display closed Design-Build-Test-Learn loops.

| Level | Description | Example |
|---|---|---|
| 5 | AI researcher | (To be achieved) |
| 4 | Highly-autonomous research | Adam, Eve [6, 7] |
| 3 | Conditional autonomy | **SDL** |
| 2 | Partial autonomy | Aquarium [5] |
| 1 | Research assistance | Pipetting robot |
| 0 | No autonomy | Electronic or paper notebook |

**Fig. 2: SDLs are level 3 autonomy systems**. Autonomy levels for SDLs describe the degree of independence from human intervention (figure adapted from Beal and Rogers [4]). At level 0, all experimental design and execution, as well as data capture, is handled by humans. At level 1, some repetitive tasks are outsourced to robots. Level 2 requires systematic digital description of protocols and experiments, as well as machine interpretable data, such as in the laboratory work planner Aquarium [5]. Level 3 involves the closed Design-Build-Test-Learn cycles that can be considered the minimum requisite for a SDL, along with interpretations of routine analyses and flagging anomalies for humans to handle. Level 4 involves robotic protocol execution and routine data analyses, as in "Adam" and "Eve" [6,7], with humans involved only as setting goals and plans (i.e., SDL works as a lab assistant to humans). At level 5, humans just set goals and receive results (i.e., SDL behaves as investigator and human as manager).

# Some examples of self-driving labs

SDLs are not unattainable fantasies: there are indeed several published examples, although focused on narrow tasks. In chemistry and materials science, SDLs are enjoying an upsurge in popularity, and several examples are now available. In biology, there are some budding examples, which show the promise of this approach.

In chemistry [3] and material sciences [2], the maturity of automation platforms and the availability of machine learning (ML) methods has enabled the creation of several SDLs or almost fully automated processes. For example, Granda *et al*. [8] developed a platform that explores the chemical space using an organic synthesis robot combined with an ML model to predict reactivity of possible reagent combinations. More recently, Christensen *et al*. [9] developed an automated closed-loop system for parallel process optimization in reactors to optimize the yield of a stereoselective Suzuki-Miyaura cross-coupling reaction. Wang *et al*. [10] developed a self-optimizing millifluidic reactor for scaling the manufacturing of nanomaterials with improved optical properties. In material sciences, Macleod *et al.* use a modular robotic platform *Ada* capable of autonomously optimizing the hole mobility of the materials commonly used in perovskite solar cells and consumer electronics [11], as well as discovering new synthesis conditions for optimized conductivities and processing temperatures for palladium films [12]. Robotics coupled with Bayesian optimization were used in multiple cases: autonomous synthesis and resistance minimization of thin films [13], optimizing mechanical properties of structures for a given application [14], improving adhesive formulations [15], achieving targeted 3D print features in additive manufacturing [16], discovering novel battery electrolytes [17], and search for photocatalyst mixtures with improved activity for hydrogen production from water [18].

Biology saw the first published closed-loop systems for scientific discovery in the form of *Adam*, a robot scientist that determined gene function through gene deletion and auxotrophic experiments in *S. cerevisiae* [6]. *Eve* followed for the repurposing of drugs, identifying an angiogenesis-inhibiting anti-cancer drug for anti-malarial use [7]. More recently, Si *et al.* [19] developed an automated platform for multiplex genome-scale engineering in *S. cerevisiae*, Hamedirad *et al.* [20] used the BioAutomata fully automated platform to optimize promoter choice in lycopene producing *E. coli*, and Kanda *et al.* [21] used an autonomous robotic system to find the optimal conditions for inducing stems cell differentiation into retinal pigment epithelial cells.

While funding for SDLs is still limited, there are instances from the US National Science Foundation (NSF), the Canadian National Research Council, and DARPA.

# The special case of biology

SDLs present unique opportunities and challenges in biology, as compared to other disciplines in which they have been deployed.

A unique opportunity is the collection of the cellular instructions in a single repository (genomic DNA) that can now be easily manipulated via gene-editing techniques. In material sciences, the Young's modulus or hole mobility of a material depend on a variety of structural and chemical elements that are distributed over the material, and can be complicated to locate and modify. In biology, a cell's phenotype is determined primarily by a combination of its environment and its genome. The genome's capability to encode an incredibly varied set of phenotypes is showcased by the fantastic diversity provided by evolution on Earth over the last three billion years. These phenotypes range from metabolic adaptation to extreme environments, carbon capture from atmosphere, production of valuable chemicals and bioproducts and multicellular coordination, to the emergence of consciousness and intelligence. Furthermore, the genome is now more accessible than ever before through recent advances in CRISPR-enabled gene editing tools [22]. This combination of accessibility, centralization and ability to produce very diverse outcomes is unparalleled and holds the promise of unique societal impact.

Conversely, distinctive hurdles involve automation capabilities that are nascent compared to other fields, and biology curricula that do not currently emphasize the backgrounds in mathematics and robotics that are critical for creating SDLs (see section "Gaps for realizing self-driving labs" for further discussion).

# Benefits of self-driving labs

The main appeal for SDLs is their ability to enable significant scientific advances, which justifies their significant cost. These scientific advances involve, firstly, solving difficult biology questions that are intractable with current approaches. Secondly, and arguably more importantly, they involve upending the development of science as we know it, to accelerate it by leveraging artificial intelligence.

The high level of investment needed to enable biological SDLs is only warranted if directed towards solving important and difficult biological problems. These involve biological problems that could take decades, or even centuries, to solve otherwise: e.g., the prediction of protein structure from sequence [23]. Some examples of remaining difficult biological problems, including both topics of fundamental and practical importance, are:

1. **Systematic increase of Titer, Rate, and Yield (TRY) for bioengineered microbial strains.** A significant obstacle in developing commercially viable processes is reaching economically viable levels of Titer, Rates and Yield of a biologically produced small molecule. The traditional approach involves heuristic combinatorial processes that rely on strain-specific in-depth metabolic knowledge (i.e. the "pull-push-block" approach [24]), and do not transfer well to other products, pathways and hosts.
2. **Mapping of regulatory networks.** Perhaps the largest hurdle in predicting an organism's metabolism is to understand how it is regulated, which involves the mechanistic understanding of a large part of its genomic complement [25].
3. **Elucidating the genotype to phenotype link.** This challenge is, arguably, the central problem in biology, but despite promising advances [26,27], it remains beyond our reach to predict accurately and quantitatively the behavior of an organism given its genome.
4. **Inverse design of microbiomes.** Microbial communities exhibit remarkable capabilities, from driving Earth's biogeochemical cycles to increasing crop productivity [28]. However, we currently lack the knowledge to design communities to meet a specification: e.g. remain stable over a year, or remove $X$ grams/liter/hour of phosphorus from wastewater.
5. **Exploring biological behavior outside Earth.** Understanding how biological systems react to being in deep space or on another planet/satellite is fundamental to enable space exploration, and the proliferation of humankind beyond a single planet. However, workforce and equipment are extremely limited in space, due to the very high logistic cost of transporting them to orbit and beyond [29].

Each of these challenges will require very different robotic setups for the corresponding SDL.

Perhaps the most important impact of SDLs in science would come from the ability to automatically build scientific knowledge. By scientific knowledge, we mean a generalized body of facts, laws, and theories able to explain and predict the behavior of the system under study. We envision SDLs to be able to draw from prior knowledge and external sources as needed to perform experiments that improve this knowledge (Fig. 3). This improvement would be reflected in increased mechanistic understanding and predictive power. We envisage that a SDL would store its accumulating knowledge as a digital twin, whose role evolves as more is learned about the biological system it is analyzing. Digital twins are virtual replicas of real-world products, systems, beings, communities, or even cities, and have become critical assets for industry [30]. The initial role, in many cases, of the digital twin would be simply to suggest experiments that identify the parts and their associations. Once sufficient experimental data have been generated to identify these associations, the role of the digital twin would be to suggest experiments that determine which correlations are causal. Once causal effects have been elucidated, the role of the digital twin would evolve into designing experiments to validate a mechanistic theory capable of explaining and quantitatively predicting these causal effects. Once this theory is calibrated, the role of the digital twin would transition into designing experiments that enable new biological systems to be built to a desired specification (inverse design). In the words of Feynman, "What I cannot create, I do not understand". We anticipate that this strategy to build scientific knowledge would involve a hybrid approach, combining pure SDLs with humans, traditional labs, and existing literature (Fig. 3).

Admittedly, this type of AI technology is not yet available, despite significant recent advances in question-answering and summarization [31], integrating prior knowledge into AI systems [32], and automated derivation of generalizable rules [33,34]. Massive language models such as GTP-3 are able to perform impressive tasks that appear to mimic natural language understanding, but these systems are ungrounded and are essentially performing pattern matching, and much needs to be done to unite classical symbolic reasoning systems with deep learning approaches [35]. Indeed, the scientific process of developing and experimentally testing hypotheses, to create a falsifiable worldview that can be used to make quantitative predictions and inform decision making comes quite close to the definition of artificial general intelligence (AGI).

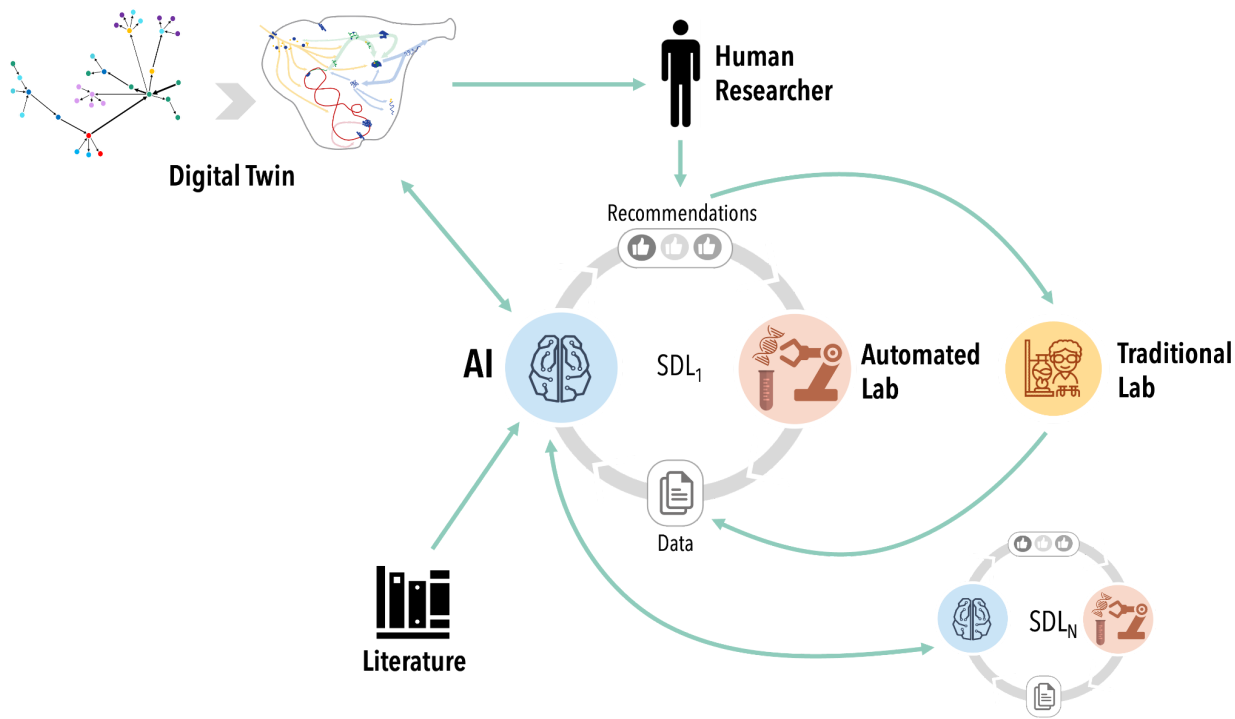

**Fig 3**: **We envision SDLs working in a network of other SDLs and humans**. To start with, in order for the SDL to make progress with respect to the current state of scientific knowledge, it must be able to draw information from existing literature. The SDL should also be able to communicate with other SDLs so as to efficiently partition the scientific phase space to be explored (e.g., $SDL_1$ focuses on one promoter set and $SDL_2$ focuses on another promoter set). Current technical limitations limit our ability to automate all experiments, so SDLs should be able

to produce unequivocal instructions for humans to follow in traditional labs, and ingest the data so produced. The final result of the operation of this network will be a digital twin of the system under study. This digital twin will likely start as a very crude and qualitative description of system parts and their connections, which will evolve as new information is obtained into more sophisticated mechanistic, quantitatively predictive models of the system under study (à la whole-cell model [26] sporting accurate predictions). These digital twins would be used by humans to access the scientific knowledge generated through this hybrid network and suggest their own recommendations. Whole-cell model figure adapted from Kerr *et al* [26].

## Gaps for realizing self-driving labs

The benefits of SDLs necessitate several technological and social advances to become a reality. These gaps involve limitations in current automation technologies, AI algorithms, data management and, importantly, sociological hurdles.

While automation of synthetic biology processes using liquid-handling commercial robotic workstations is gaining momentum, this approach has limitations for SDLs which new technologies may help ease. Companies such as, e.g., Ginkgo Bioworks or Amyris automate their discovery process using these workstations, and a few are even providing automation as a service [36]. However, the processes automated in the chemistry and material sciences SDLs discussed above are only a subset of the ones needed in synthetic biology. Typical molecular biology processes like cell transformations via electroporation, colony picking, plating and outgrowth, while doable through liquid handlers and other instrumentation, are very difficult to link together in the seamless manner SDLs require (Fig. 2). Microfluidics offer the opportunity to provide this seamless integration by encapsulating cells and reagents into droplets, and manipulating them precisely. Indeed, microfluidic platforms have been proposed for miniaturization of biological reactions including DNA synthesis [37], DNA assembly [38], transformation [39], cell-free expression [40], and phenotypic screening by fluorescence [41] and mass spectrometry [42]. Truly disruptive functionalities can be achieved by combining these capabilities with new developments in molecular sensors embedded on semiconductor chips [43], wireless optically-activated microscopic sensors [44], monitoring of free radicals through fluorescent nanodiamonds [45], metabolic modulation through optogenetics [46], or manipulation of cells with light [47]. Microfluidic sampling from bioreactors can also enable real-time sensing and imaging of cells in their environments, enabling continuous data capture. Moreover, these microfluidic platforms are far more affordable and use less reagents than robotic workstations, permitting a much larger number of experiments and democratizing the access to synthetic biology. Their routine use in synthetic biology, however, necessitates sustained investment to enable seamless functioning and the automation of the full range of synthetic biology processes.

Novel AI algorithms are needed to make SDLs a reality in synthetic biology. Although current algorithms can guide the metabolic engineering process effectively [48], wide-spread adoption of SDLs will require the AI to understand context, and the ability to produce interpretable knowledge. This means the ability to: 1) use prior knowledge to inform the AI in the SDL, and 2) extract knowledge out of the predictive capabilities of the AI such that it can be extrapolated to related, but different, experimental conditions by other human researchers or SDLs (Fig. 3). The ability to leverage and produce extrapolatable knowledge is critical if we are to benefit from a large amount of SDLs. Otherwise, humans would become the bottleneck in transferring the knowledge accumulated in the digital twins from and to the SDLs (Fig. 3). One possibility to introduce this much needed context may lie in the use of foundational models [49], trained on massive datasets, and adapted to specific use cases.

Data management is a critical link between automation and AI algorithms that has been often neglected in the past. While often considered a burdensome chore, there is simply no AI without data, and there are no self driving labs without AI. General ontologies and extensible standards for data and protocols are critical if large amounts of data are to be collected and seamlessly integrated into an ecosystem involving continuous data exchange among SDLs and human researchers.

Another important obstacle for the creation of SDLs in biology involves the sociological challenges in having computer scientists, molecular biologists and automation engineers work together. These two worlds embody very different scientific cultures, which are reflected not only in how they solve problems, but also which problems they consider worth solving [50]. Having them work together constructively is, arguably, harder than the technological challenges faced by SDLs in biology. Currently, computational and bench scientists are trained very differently; a critical first step is to design a training curriculum that exposes them to each other's world.

# Conclusion

While SDLs are bound to be costly endeavors, the expected returns make them worthwhile undertakings. A fully functioning network of SDLs and human researchers (Fig. 3) would not only provide significant biological knowledge, but also the ability to fully exploit synthetic biology for biomanufacturing purposes. Furthermore, they would provide the opportunity to understand and improve the process of constructing scientific knowledge. In that sense, the large project of creating SDLs mirrors the Human Genome Project, in that they show a potential to fundamentally transform the field of biology.

We must, however, be aware of the risks associated with SDLs: their use for nefarious purposes (e.g. virus synthesis), including the ability to be manipulated via remote cyberattacks. A more subtle risk involves the possible long-term misalignment with our values and goals, which can be challenging to fully encode in a machine readable manner, potentially allowing the system to act in an unintended or undesired manner.

Despite the risks and challenges, we believe that SDLs represent the next leap forward in the progress of scientific research, and that synthetic biology poses a unique opportunity for their development.

# Acknowledgments

This work was part of the DOE Agile BioFoundry (http://agilebiofoundry.org), the Advanced Biofuels and Bioproducts Process Development Unit (https://abpdu.lbl.gov/), and the DOE Joint BioEnergy Institute (http://www.jbei.org), supported by the U. S. Department of Energy, Energy Efficiency and Renewable Energy, Bioenergy Technologies Office, and the Office of Science, through contract DE-AC02-05CH11231 between Lawrence Berkeley National Laboratory and the U.S. Department of Energy. S.P. and D.A. were supported by Laboratory Directed Research and Development (LDRD) funds provided by Lawrence Berkeley National Laboratory, operated for the U.S. Department of Energy under the same contract. HGM was also supported by the Basque Government through the BERC 2018–2021 program and by the Spanish Ministry of Economy and Competitiveness MINECO: BCAM Severo Ochoa excellence accreditation SEV-2017-0718. K.E.B. was funded by the Department of Energy, Advanced Scientific Computing Research. J.M.M. was supported by the U.S. Department of Energy (DOE), Office of Science, Office of Biological and Environmental Research, Lawrence Livermore National Laboratory (LLNL) SFA "From Sequence to Cell to Population: Secure and Robust Biosystems Design for Environmental Microorganisms," under Contract DE-AC52-07NA27344 (LLNL-JRNL-837127). J.M.C. was supported in part by the U. S. Department of Energy, Energy Efficiency and Renewable Energy, Bioenergy Technologies Office under contract DE-EE0008927. Gy.B. was supported by the "Rapid Design and Engineering of Smart and Secure Microbiological Systems" project funded by the Biological Systems Science Division's Genomic Science Program, within the U.S. Department of Energy, Office of Science, Biological and Environmental Research. Argonne National Laboratory is managed by UChicago Argonne, LLC for DOE under contract number DE-AC02-06CH11357. LW was funded by the US National Science Foundation Graduate Research Fellowship. The views and opinions of the authors expressed herein do not necessarily state or reflect those of the United States Government or any agency thereof. Neither the United States Government nor any agency thereof, nor any of their employees, makes any warranty, expressed or implied, or assumes any legal liability or responsibility or the accuracy, completeness, or usefulness of any information, apparatus, product, or process disclosed or represents that its use would not infringe privately owned rights.

The United States Government retains and the publisher, by accepting the article for publication, acknowledges that the United States Government retains a nonexclusive, paid-up, irrevocable, worldwide license to publish or reproduce the published form of this manuscript, or allow others to do so, for United States Government purposes. The Department of Energy will provide public access to these results of federally sponsored research in accordance with the DOE Public Access Plan (http://energy.gov/downloads/doe-public-access-plan). Funding for open access charge: US Department of Energy.

# Annotated references

## Outstanding interest:

[8] (Granda et al): A great example of the potential of SDLs, showing how a robot is able to systematically explore chemical space and successfully predict reactivity.

[18] (Burger et al): An inspiring use of a mobile robotic arm to automate the researcher rather than the instruments, opening the transition to SDLs for any traditional lab.

[32] (Cai et al): An informative review on how to embed prior knowledge in AI, in this case for fluid dynamics in the form of PINNs (Physics-Informed Neural Networks). Similar approaches would be needed for biology.

[34] (Guimera et al): A stimulating demonstration of the power of "machine scientists", able to extract closed mathematical models automatically out of data.

[47] (Rienzo et al): This paper demonstrates the use of microfluidics and automated cell manipulation through light for synthetic biology, providing a promising platform for SDLs.

[43] (Fuller et al): A very interesting report on the possibilities created by embedding single molecules in electronic chips.

## Special interest:

[6] (King et al): *Adam* is the first published example of a closed-loop system that designs and executes experiments to test inferred hypotheses. A classic well before SDLs became of widespread interest.

[7] (Williams et al): *Eve* constitutes an outstanding example of the use of SDLs to alleviate the large cost of drug discovery.

[26] (Karr et al): One of the first examples of a whole-cell model, accounting for all annotated gene functions in *Mycoplasma genitalium*, and validated against a broad

range of data.

[30] (Jian et al): Good introduction to digital twins, and how they are becoming an industry staple.

[38] (Gach et al): A nice demonstration of what microfluidics can achieve in terms of automating synthetic biology protocols.

[48] (Lawson et al): Interesting review on the current and possible applications of AI in metabolic engineering and synthetic biology.